# $^{59}$Co-NMR Knight Shift of Superconducting Three-Layer $Na_xCoO_2 \cdot yH_2O$

H. Watanabe, Y. Kobayashi and M. Sato*

*Department of Physics, Division of Material Science, Nagoya University, Furo-cho, Chikusa-ku, Nagoya 464-8602*

The superconducting state of $Na_xCoO_2 \cdot yH_2O$ with three $CoO_2$ layers in a unit cell has been studied by $^{59}$Co-NMR. The Knight shift measured for a peak of the NMR spectra corresponding to the external magnetic field $\boldsymbol{H}$ along one of the principal directions within the $CoO_2$ plane, exhibits a rapid decrease with decreasing temperature $T$ below the superconducting transition temperature $T_c$, indicating that the spin susceptibility $\chi_{spin}$ is suppressed in the superconducting phase, at least, for this field direction. Because differences of the superconducting properties are rather small between this three-layer $Na_xCoO_2 \cdot yH_2O$ and previously reported $Na_xCoO_2 \cdot yH_2O$ with two $CoO_2$ layers within a unit cell, the present result of the Knight shift studies indicates that the Cooper pairs of the former system are in the singlet state as in the latter, for which the spin susceptibility is suppressed for both directions of $\boldsymbol{H}$ parallel and perpendicular to the $CoO_2$ plane.

## 1. Introduction

The hydrated Co oxide $Na_xCoO_2 \cdot yH_2O$ ($x$~0.3 and $y$~1.3) having two $CoO_2$ layers within a unit cell, which are formed of edge-sharing $CoO_6$ octahedra exhibits the superconducting transition at temperature $T = T_c$~4.5 K.[1] Because the layers have the triangular lattice of Co atoms, roles of the geometrical frustration are, naively speaking, expected to be important for the occurrence of the superconductivity, and much attention has been paid to the system. It is also presumed that strong spin excitations are related to the superconductivity, because the spin state change is often found in $Co^{3+}$ ions in various kinds of Co oxides.[2-4] Then possible exotic characters have been intensively pursued and the pairing symmetry is being hotly argued.[5-12]

We have studied the $^{59}$Co-NMR Knight shift $K$ of two-layer $Na_xCoO_2 \cdot yH_2O$[6,7] and reported that it decreases with decreasing $T$ below $T_c$ for both directions of the applied field $\boldsymbol{H}$ within the *ab*-plane and parallel to the *c*-axis.[7] The result indicates that the superconducting electron pairs are in the singlet state.

In this paper, results of the $^{59}$Co-NMR studies on another superconducting system of $Na_xCoO_2 \cdot yH_2O$,[13-15] which has three layers of edge-sharing $CoO_6$ octahedra within a unit cell, are presented. The $T_c$ value of this three-layer $Na_xCoO_2 \cdot yH_2O$ is ~4.5 K, nearly equal to that of two-layer $Na_xCoO_2 \cdot yH_2O$.

In Fig. 1, the structures of (a) P2-$Na_{0.75}CoO_2$ and (b) O3-$NaCoO_2$[16,17] are compared. In the figure, we can recognize that the repetition of the layer sequences of ($O_A$-Co-$O_B$-Na-$O_B$-Co-$O_A$-Na-) and ($O_{A'}$-Co-$O_{B'}$-Na-$O_{C'}$-Co-$O_{A'}$-Na-$O_{B'}$-Co-$O_{C'}$-Na-) within a unit cell are repeated, for P2-$Na_{0.75}CoO_2$ and (b) O3-$NaCoO_2$, respectively, where $O_X$ (X=A, B, A', B' and C') distinguish the oxygen layers. The superconducting systems are prepared by de-intercalating Na atoms and then, intercalating $H_2O$ molecules. During these processes, the stacking sequence of P2-$Na_xCoO_2$ is kept unchanged, while the O3-$NaCoO_2$ is transformed to the P3 type one with a sequence of (-$O_{A''}$-Co-$O_{B''}$-Na-$O_{B''}$-Co-$O_{C''}$-Na-$O_{C''}$-Co-$O_{A''}$-Na-).[15]

We have performed $^{59}$Co-NMR studies of the powder samples of P3-$Na_xCoO_2 \cdot yH_2O$ and compared the observed results with those of P2-$Na_xCoO_2 \cdot yH_2O$.

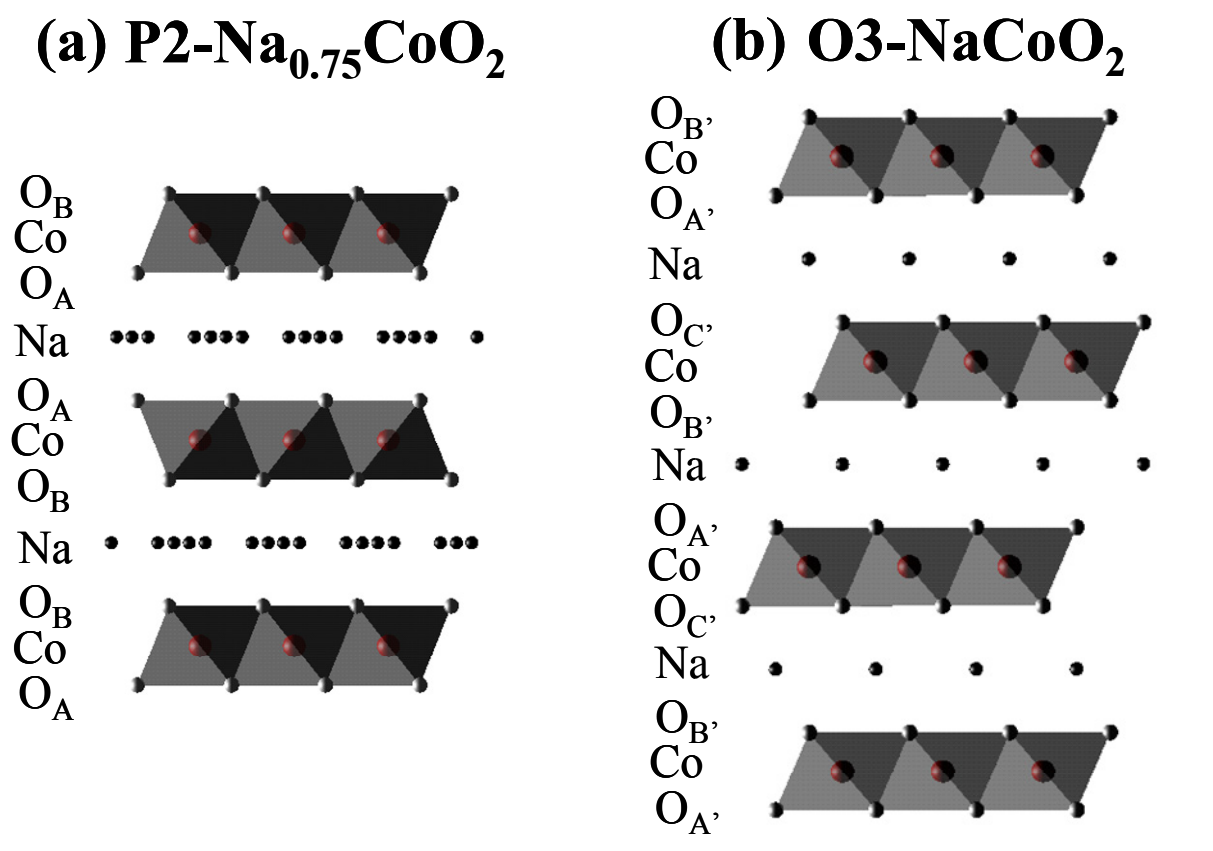

Fig. 1. Crystal structures of (a) P2-$Na_{0.75}CoO_2$ and (b) O3-$NaCoO_2$, where the Na sites are shown by the dark spheres and Co sites are within the shaded octahedra formed of the corner oxygens. In O3-$NaCoO_2$, the Na sites are fully occupied, while in P2-$Na_{0.75}CoO_2$, there are two crystallographically distinct Na sites with different occupancies.

## 2. Syntheses and Characterizations of the samples

Powder samples of P3-$Na_xCoO_2 \cdot yH_2O$ were synthesized by using the procedure described in ref. 13: The mother system O3-$NaCoO_2$ was synthesized, by heating NaOH pellets buried in the powder of Co metal in air at 500 °C for 12 h, where we had to use the 25 % excess amount of NaOH to obtain O3-phase samples of $NaCoO_2$. (If the molar ratio of Na and Co is smaller than ~1.25, a significant amount of P2-$Na_xCoO_2$ with $x$~0.75 is found as the impurity phase.) The fused ingots were finely ground and annealed at 800 °C under a $N_2$ flow for 16 h. In the X-ray diffraction pattern of the sample obtained by this heat treatments, we did not see any impurity phases. It is noted that the obtained O3-$NaCoO_2$ tends to change gradually to P2-$Na_{0.75}CoO_2$ in air. The powder was immersed in the $Br_2/CH_3CN$ solution for 5 days to de-intercalate Na atoms and then the product was washed with $CH_3CN$ and dried in air. Finally, in order to obtain the superconducting hydrated phase with two $H_2O$ layers between the neighboring $CoO_2$ planes (bilayer phase), the product was sealed in a chamber with saturated $H_2O$ vapor at room temperature for a week.

*corresponding author: (e-mail: e43247a@nucc.cc.nagoya-u.ac.jp)

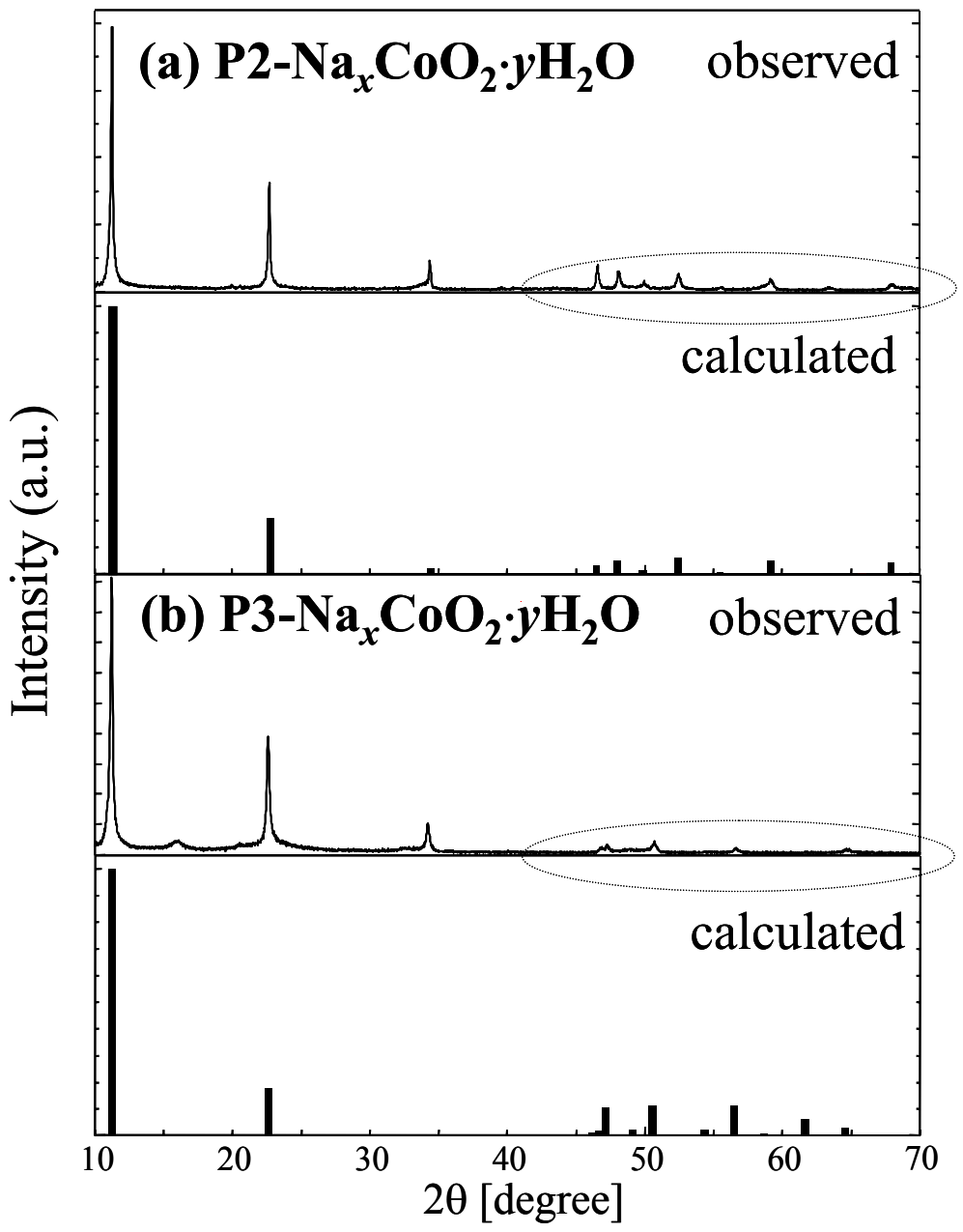

Fig. 2. The X-ray powder diffraction patterns observed for the samples of (a) P2- and (b) P3-$Na_xCoO_2{\cdot}yH_2O$ are compared with the integrated intensities calculated by using the space groups $P6_3/mmc$ and $R\bar{3}m$, respectively. The data in the high angle regions surrounded by the broken lines indicate the clear distinction between the structures of P2- and P3-$Na_xCoO_2{\cdot}yH_2O$.

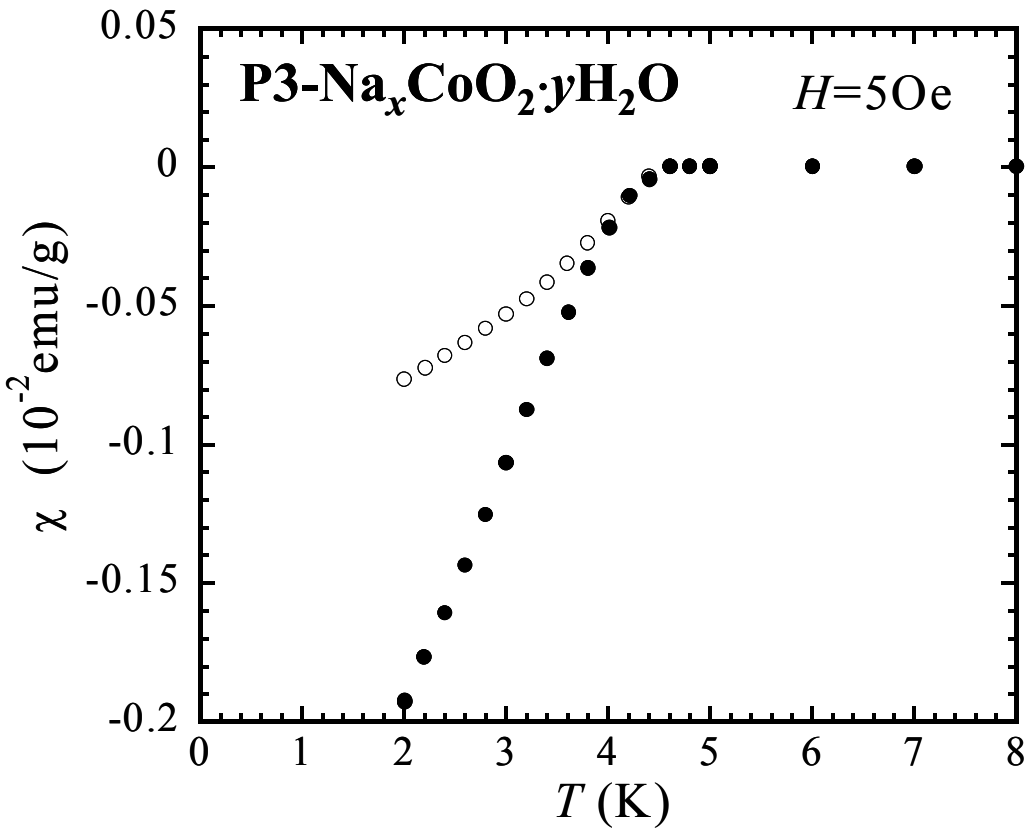

Fig. 3. Magnetic susceptibility of P3-$Na_xCoO_2{\cdot}yH_2O$ measured with the magnetic field $H$ = 5 Oe. Open and solid circles are measured under the conditions of the field cooling and the zero field cooling, respectively.

The obtained powder sample of P3-$Na_xCoO_2{\cdot}yH_2O$ was characterized as a single phase by the powder X-ray diffraction. In Figs. 2(a) and 2(b), the observed X-ray diffraction patterns of P2- and P3-$Na_xCoO_2{\cdot}yH_2O$ are compared with the calculated ones, respectively (space group $P6_3/mmc$ for the former and $R\bar{3}m$ for the latter). The preparation method of the P2-$Na_xCoO_2{\cdot}yH_2O$ can be found elsewhere.[6,7] We can distinguish these two phases from the peak positions in the high angle region. The lattice parameters for the samples of P3-$Na_xCoO_2{\cdot}yH_2O$ are $a = b = 2.8253(7)$ and $c = 29.423(6)$ Å, which are consistent with previously reported ones[13-15].

Figure 3 shows the $T$ dependence of the magnetic susceptibility measured with the magnetic field $H$ = 5 Oe for a powder sample of P3-$Na_xCoO_2{\cdot}yH_2O$ used in the NMR/NQR measurements. The superconducting transition temperature $T_c$ is about 4.5 K, which is nearly equal to that of P2-$Na_xCoO_2{\cdot}yH_2O$.

$^{59}$Co-NMR spectra were obtained for the randomly oriented powder sample of P3-$Na_xCoO_2{\cdot}yH_2O$ by the spin-echo technique using a phase coherent type pulse spectrometer. The echo intensity was recorded with the applied magnetic field changed stepwise. The nuclear longitudinal relaxation rate $1/T_1$ of $^{59}$Co nuclei was measured by measuring the $^{59}$Co nuclear magnetization $m$ as a function of the time $t$ elapsed after applying an inversion pulse in the frequency region around $3\nu_Q$, $\nu_Q$ being the electrical quadrupolar frequency (~4 MHz). The $\{1-m(t)/m(\infty)\}-t$ curves were found to be described by the theoretical one. Other details can be found in ref. 6.

### 3. Results of the NMR Studies and Discussion

Field swept spectra of $^{59}$Co NMR taken at 5 K with the frequency $f$ = 23.566 MHz are shown in Fig. 4 together with the result of the model calculation (broken line) obtained as described in ref. 6 for P2-$Na_xCoO_2{\cdot}yH_2O$. In the present case, the used NMR parameters are as follows. The quadrupolar frequency $\nu_Q$ = 4.1 MHz, the Knight shifts $K_x$ = 3.2 %, $K_y$ = 3.5 % and $K_z$ = 2.1 % for locally defined principal axes at Co sites, $x$, $y$ and $z$, respectively and the asymmetric parameter $\eta$ = 0.3. These parameters, which can well reproduce the observed spectra, are rather similar to those obtained for P2-$Na_xCoO_2{\cdot}yH_2O$.[6)] Because the trigonal distortions of $CoO_6$ octahedra in P2- and P3-$Na_xCoO_2{\cdot}yH_2O$ are almost equal,[18)] we can expect that the principal axis $z$ of the latter system is parallel to the $c$ axis, as in P2-$Na_xCoO_2{\cdot}yH_2O$. The peak indicated by the arrow in Fig. 4 corresponds to the magnetic field direction parallel to one of the principal directions ($y$-direction) within the $ab$-plane. We have measured the $T$-dependence of this peak at frequencies $f$ of 23.566, 33.281 and 43.625 MHz. The value of the resonance field $H_{res}$ determined as the peak position of the magnetic field $H$ begins to exhibit rather rapid increase with decreasing $T$ at the transition temperature $T_c$ for all frequencies. This behavior is similar to that of P2-$Na_xCoO_2{\cdot}yH_2O$.[6,7)] For example, the peak profiles observed at various temperatures for $f$ = 23.566 MHz are shown in Fig. 5, where the peak values are adjusted to that shown in Fig. 4 (in Fig. 5, almost the whole part of the profiles above the dip found at the higher field side of the peak is shown and the shift of the peak toward the higher field side with decreasing $T$ is evident). The shift occurs due to the magnetic shift and/or the effect of the second order and higher order $eqQ$ interactions. In the present case, it is considered that the magnetic shift predominantly contributes to the observed one, because the $T$ dependence of the shift originating from the quadrupolar interaction is considered to be negligibly small in this low temperature region.

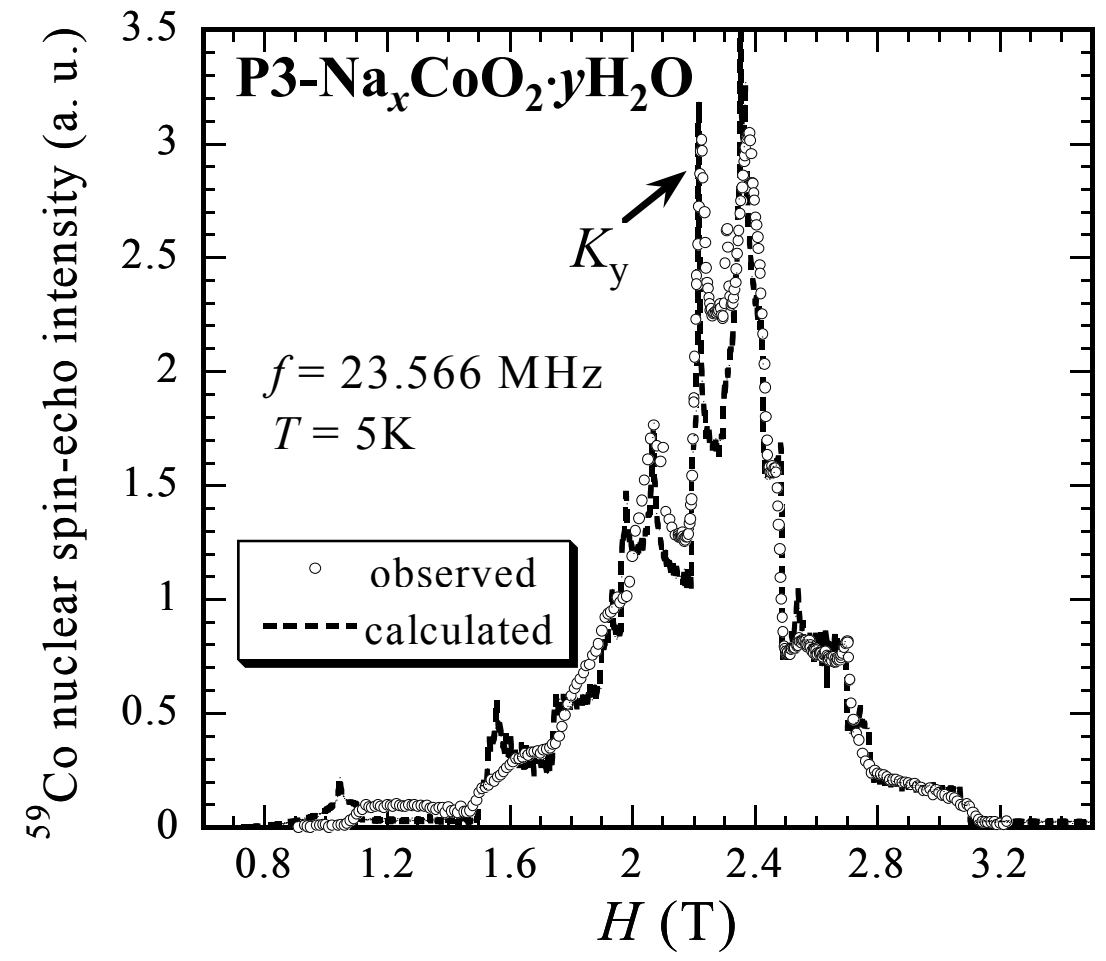

Fig. 4. Field swept $^{59}$Co-NMR spectra of P3-$Na_xCoO_2$·$yH_2O$ taken at 5 K with a fixed frequency of 23.566 MHz for randomly oriented powder samples. The broken line shows the powder pattern calculated by the parameters $\nu_Q$ = 4.1 MHz, $K_x$ = 3.2 %, $K_y$ = 3.5 %, $K_z$ = 2.1 % and $\eta$ = 0.3.

In Fig. 6(a), the deviation of the resonance field $H_{res}$ from the value just above $T_c$, $\Delta H_{res}(T) \equiv H_{res}(T)-H_{res}(T>T_c)$ observed for various frequencies (or various applied field) are shown. It is worth noting that at low temperatures the absolute value of $\Delta H_{res}$ increases with increasing $H$, which was also observed for P2-$Na_xCoO_2$·$yH_2O$. This field dependence of $\Delta H_{res}$ cannot be explained solely by the superconducting diamagnetism $M_{dia}$, because the absolute value of $M_{dia}$ decreases with increasing $H$ in the measured $H$ region ($H >> $ the lower critical field $H_{c1}$). Consequently, we can say that for the nonzero $\Delta H_{res}$ observed below $T_c$, the reduction of the spin magnetization $M_{spin}$ is important. (The orbital contribution to the Knight shift can be considered to be $T$ independent in this low temperature region.)

Figure 6(b) shows the $T$-dependence of $\Delta K(T) \equiv f/\{\gamma_N H_{res}(T)\}-f/\{\gamma_N H_{res}(T>T_c)\}$ for various fixed frequencies, where $\gamma_N$ is the nuclear gyromagnetic ratio ($\gamma_N$ = 10.03 MHz/T for $^{59}$Co). If we neglect the contribution of $M_{dia}$, the reduction of $\Delta K$ is due to the suppression of the spin component $K_{spin}$ of the Knight shift $K$. The magnitudes of $\Delta K$ for the present P3-$Na_xCoO_2$·$yH_2O$ are slightly smaller than those for P2-$Na_xCoO_2$·$yH_2O$.[6,7] But, in both systems, $K_{spin}$ decreases below $T_c$, suggesting that the superconducting electron pairs should be considered to be in the singlet state, unless the spins of Cooper pairs in the triplet state are strongly pinned along the $c$-direction. On this point, we have already reported[7] that in P2-$Na_xCoO_2{\cdot}yH_2O$, $\Delta K$ decreases (or $|\Delta K|$ increases) with decreasing $T$ below $T_c$ for both directions of $\boldsymbol{H}$ within the $ab$-plane and parallel to the $c$-axis. It indicates that the pairs are in the singlet state in P2-$Na_xCoO_2{\cdot}yH_2O$.

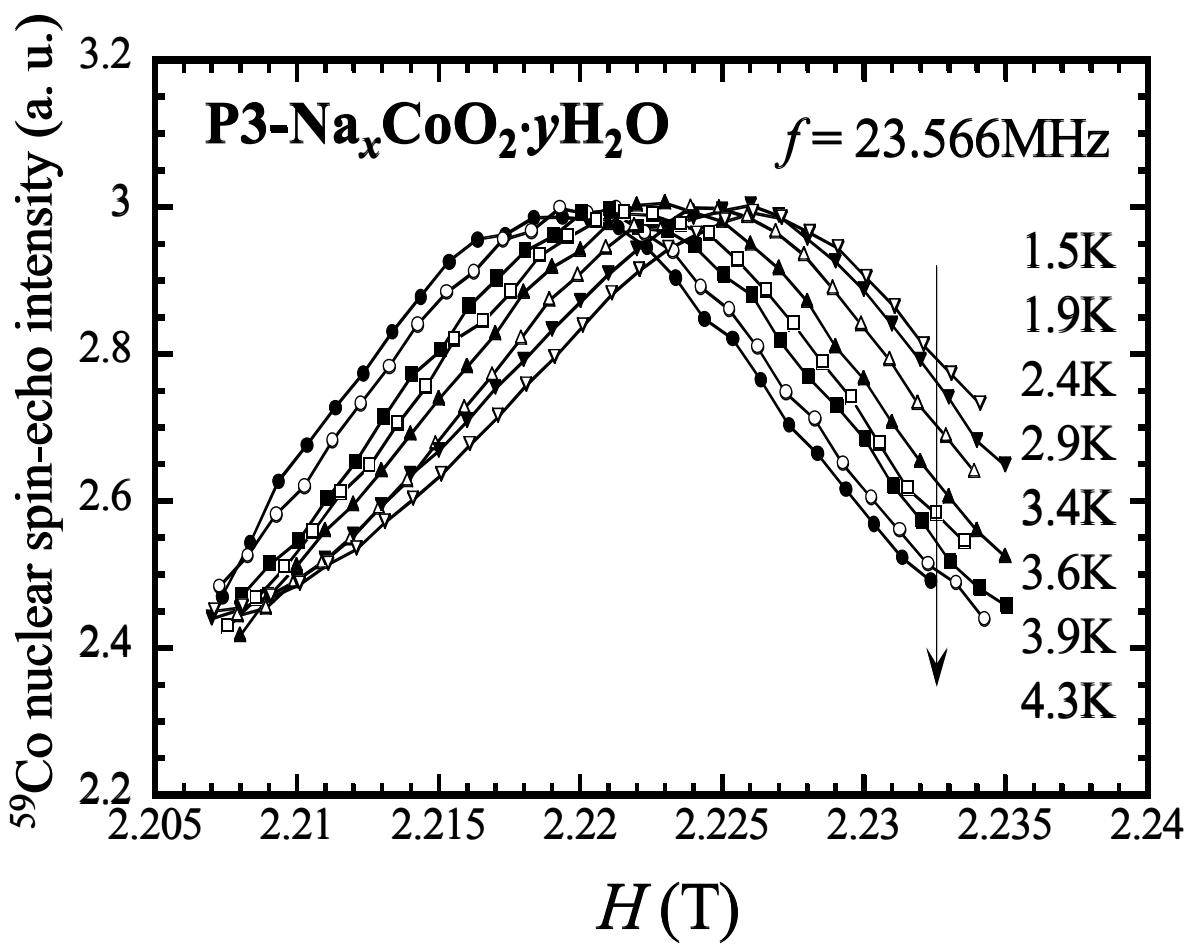

Fig. 5. Temperature dependence of the profile of the peak indicated in Fig. 4 by the arrow is shown for a fixed frequency of 23.566MHz. The solid lines are the guides for eyes.

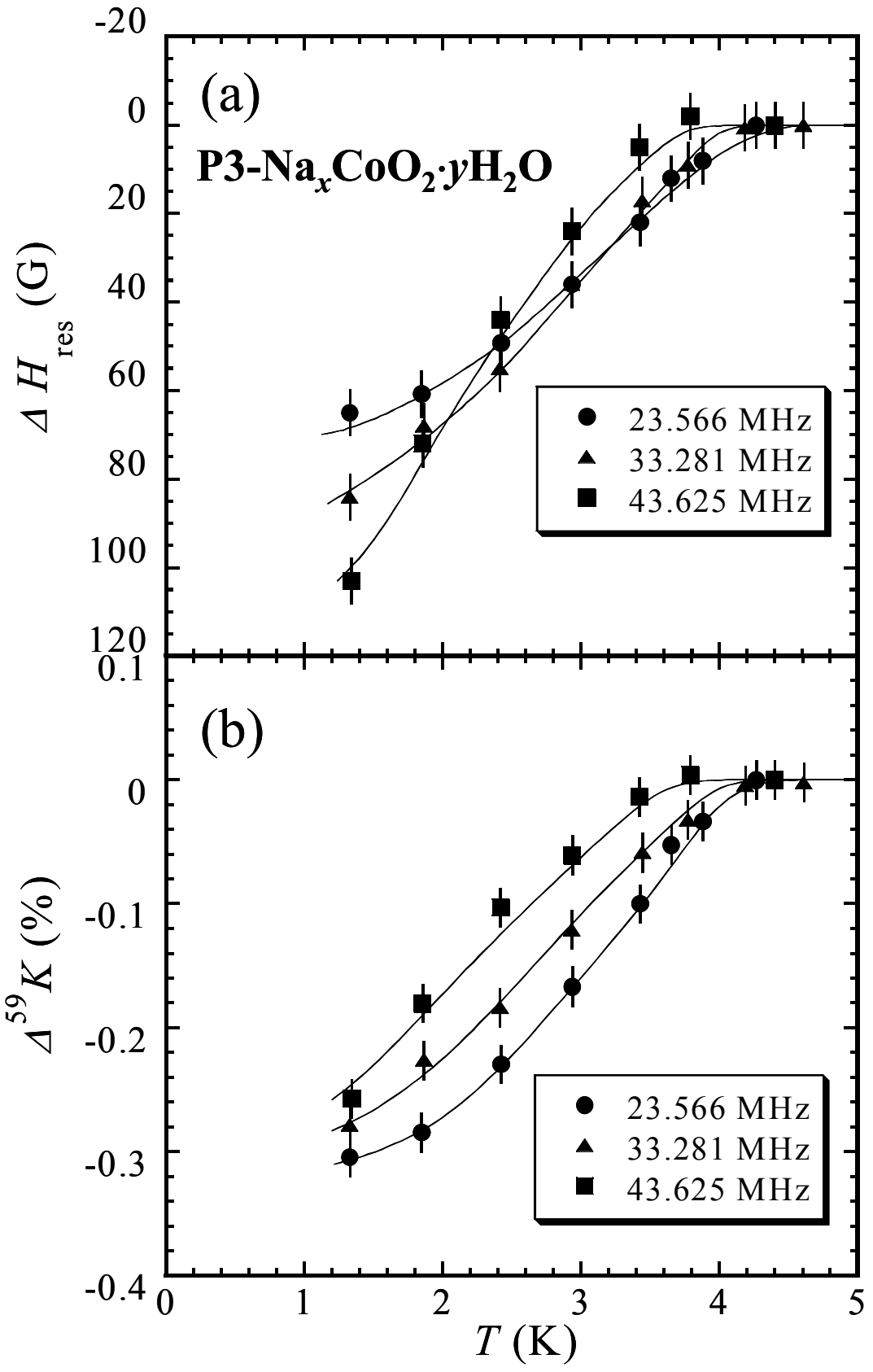

Fig. 6. (a) Deviations of the resonance field from the value just above $T_c$, $\Delta H_{res}(T) \equiv H_{res}(T)-H_{res}(T>T_c)$ are plotted against $T$ for $\boldsymbol{H}$ // $y$ at three fixed frequencies. (b) Deviations of the Knight shift from the value just above $T_c$, $\Delta K(T) \equiv f/\{\gamma_N H_{res}(T)\}-f/\{\gamma_N H_{res}(T>T_c)\}$ are plotted against $T$ for various fixed frequencies, where $\gamma_N$ is the nuclear gyromagnetic ratio ($\gamma_N$ = 10.03 MHz/T for $^{59}$Co).

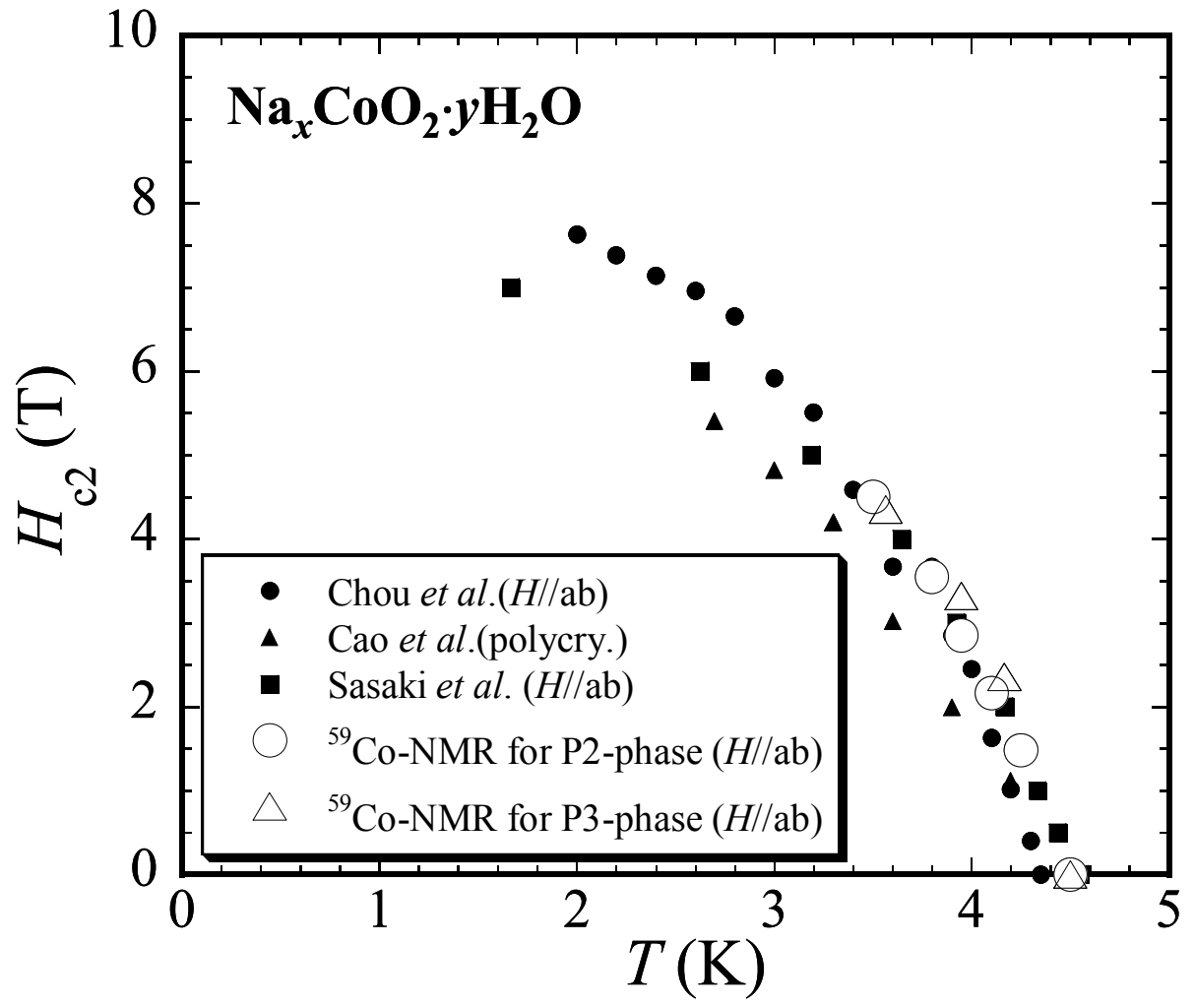

Fig. 7. $T$-dependence of the upper critical field $H_{c2}$ for $\boldsymbol{H}$ // $ab$-plane determined from the $\Delta K(H)$-$T$ curves of P2- and P3-$Na_xCoO_2{\cdot}yH_2O$ is shown together with the reported data of the resistivity measurements.

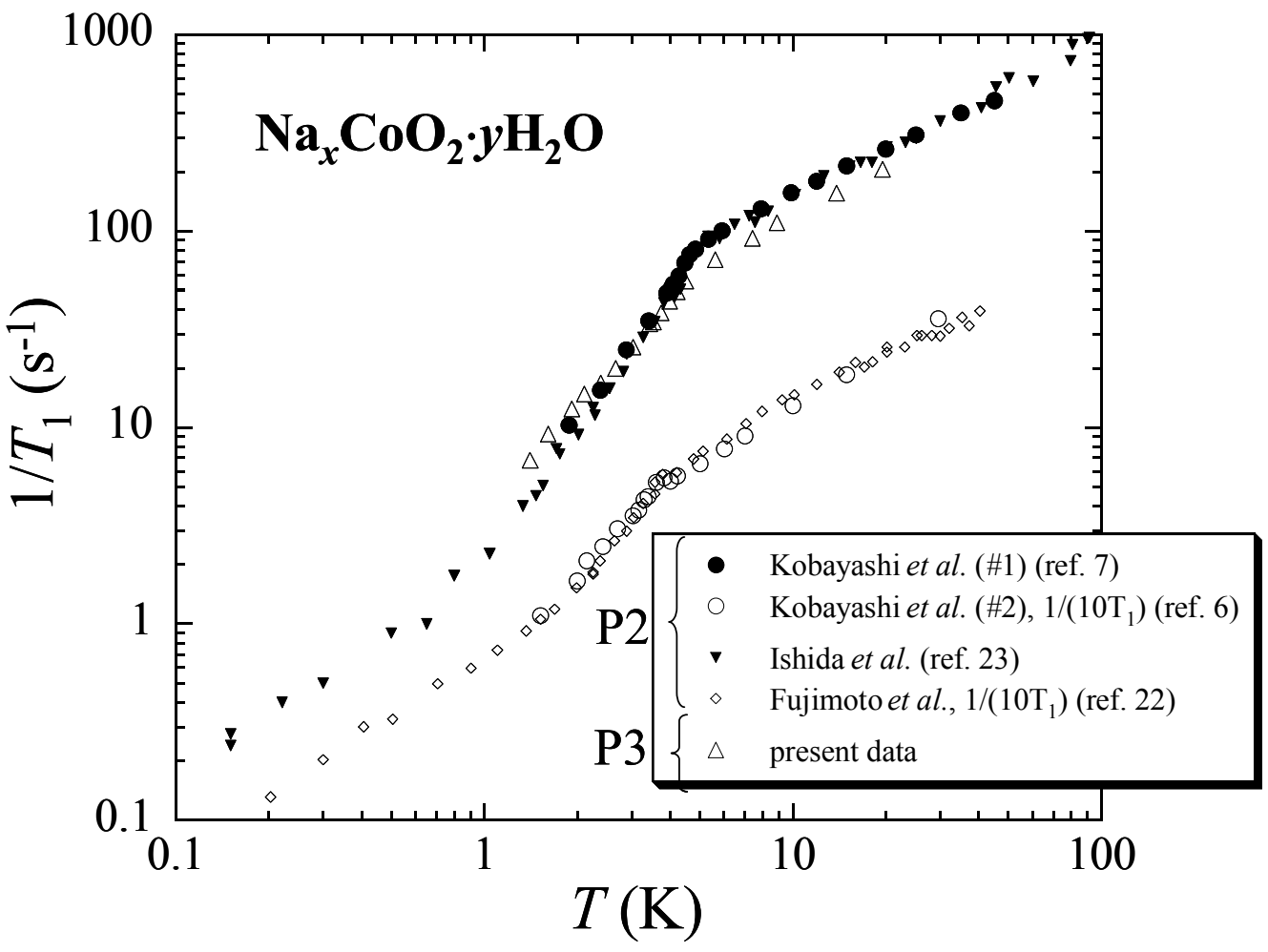

Fig. 8. Longitudinal relaxation rate $1/T_1$ of $^{59}$Co nuclei are plotted against $T$ in logarithmic scales. Several data reported for P2-$Na_xCoO_2{\cdot}yH_2O$ are also shown.

Figure 7 shows the $T$-dependence of the upper critical field $H_{c2}$ estimated from $\Delta K(H)$-$T$ curves for P3- and P2-$Na_xCoO_2{\cdot}yH_2O$, together with those of resistivity measurements for P2-$Na_xCoO_2{\cdot}yH_2O$.[7,19-21] The value of $H_{c2}(T\to 0)$ is estimated to be about 8 T, which well agrees with the value expected from the Pauli paramagnetic limit $H_p = 1.84T_c$ for systems with the negligible effect of the spin-orbit coupling and with $T_c$ = 4.5 K. Because the Pauli limit arises from the pair breaking by the Zeeman effect, the observed value of $H_{c2}$ implies that the singlet superconductivity is realized in the present cobalt oxide superconductor P3-$Na_xCoO_2{\cdot}yH_2O$. Thus, we may consider that both of P2- and P3-$Na_xCoO_2{\cdot}yH_2O$ possess the same superconducting pair symmetry. For P2-$Na_xCoO_2{\cdot}yH_2O$, we have shown from the measurements of the Knight shifts that the spin susceptibility $\chi_{spin}$ is suppressed for both directions of $\boldsymbol{H}$ within and perpendicular to the $ab$-plane.[7] For P3-$Na_xCoO_2{\cdot}yH_2O$, we have not studied the shift for the perpendicular field, because it is difficult to grow single crystals of the mother system O3-$NaCoO_2$. However, the similarity of the data shown in the present paper for $\boldsymbol{H}$ within the $ab$-plane implies that no significant difference exists between the behaviors of the Knight shifts of P2- and P3- superconductors. Then, we think that the singlet pairing is also realized in the present P3-$Na_xCoO_2{\cdot}yH_2O$.

In Fig. 8, the longitudinal relaxation rate, $1/T_1$ is plotted against $T$ for $^{59}$Co nuclei. In the figure, the data reported for P2-$Na_xCoO_2{\cdot}yH_2O$[6,7,22,23] are also shown. We find that both the P2- and P3-$Na_xCoO_2{\cdot}yH_2O$ systems exhibit similar $T$ dependences of $1/T_1$ and that they do not exhibit the coherence peak, confirming that their superconducting phases exhibit very similar behaviors. As has been already discussed in the preceding paper,[7] it seems to be important, for the complete understanding of the superconducting state of the hydrated Co oxides, to have a consistent explanation of the behaviors of $K$, $1/T_1$ and the rather small effect of non-magnetic impurities on the superconducting transition temperature $T_c$.[8] It remains as a future problem.

Neutron scattering studies carried out for P2-$Na_{0.82}CoO_2$[24] and P2-$Na_{0.75}CoO_2$[25] have shown that the interlayer magnetic interaction of the electron system is so significant that the system has the 3-dimensional nature. It is also reported that the in-plane magnetic correlation is ferromagnetic for these Na-rich systems. It has been pointed out that the origin of this significant interlayer coupling takes place through the $sp^2$ hybridized orbits of $Na^+$ and $O^{2-}$ ions between the $CoO_2$ planes.[26] Then, the neutron results stated above might indicate that the stacking way of the $CoO_2$ layers, which affects the interlayer coupling, is important for the determination of the physical properties of $Na_xCoO_2$. However, various kinds of results obtained by NMR/NQR, neutron scattering and other studies on P2-$Na_xCoO_2$ indicate that the system has the low dimensional nature and the antiferromagnetic in-plane correlation of the electrons in the region of $x \leq 0.6$, as reported elsewhere.[27] Moreover, it is expected for the superconducting systems with $x \sim 0.3$, the 2-dimensional nature is enhanced by the $H_2O$ intercalation. Therefore, we think that the stacking way of the $CoO_2$ planes does not have effects on the physical properties. It is consistent with the present experimental results.

## 4. Summary

We have presented the results of the $^{59}$Co-NMR experiments for a randomly oriented sample of P3-$Na_xCoO_2{\cdot}yH_2O$. The magnitude of the Knight shift $K$ decreases with decreasing $T$ below $T_c$ for the field direction $\boldsymbol{H}$ within the $ab$-plane. For this reduction, the suppression of $K_{spin}$ is essential. The value of $H_{c2}(T\to 0)$ estimated from the $\Delta K(H)$-$T$ curves is similar to those of P2-$Na_xCoO_2\cdot yH_2O$. The structural difference between P3- and P2-$Na_xCoO_2\cdot yH_2O$ seems not to induce any significant differences between the superconducting properties of these two systems. In the strict sense, the present results do not completely exclude the possibility of the triplet pairing state with their spins strongly pinned along the *c*-axis. However, considering that $\Delta K_{spin}$ of P2-$Na_xCoO_2\cdot yH_2O$ decreases with decreasing $T$ below $T_c$ even in the case of $\boldsymbol{H}$ parallel to the *c* direction, it is natural to conclude that the possibility of the triplet pairing is ruled out for the present P3-$Na_xCoO_2\cdot yH_2O$, too.

**Acknowledgements**

This work is supported by Grants-in-Aid for Scientific Research from the Japan Society for the Promotion of Science (JSPS) and by Grants-in-aid on priority areas from the Ministry of Education, Culture, Sports, Science and Technology.